\documentclass[preprint]{revtex4}
\usepackage{epsfig}

\def\beq{\begin{equation}}
\def\eeq{\end{equation}}
\def\beqa{\begin{eqnarray}}
\def\eeqa{\end{eqnarray}}

\begin{document}

\title{\sc Solitons in relativistic mean field models}

\author{D.A. Foga\c{c}a\dag\  and F.S. Navarra\dag\ }
\address{\dag\ Instituto de F\'{\i}sica, Universidade de S\~{a}o Paulo\\
 C.P. 66318,  05315-970 S\~{a}o Paulo, SP, Brazil}

\begin{abstract}
Assuming that the nucleus can be treated as a perfect fluid we study the conditions for 
the formation and propagation
of Korteweg-de Vries (KdV) solitons in nuclear matter.  The KdV equation
is obtained from the  Euler  and continuity equations in 
nonrelativistic hydrodynamics.  The existence of these solitons depends on the nuclear 
equation of state, which, in our approach,  comes from well known relativistic mean field 
models. We reexamine early works 
on nuclear solitons, replacing the old equations of state by new ones,  based on QHD and on 
its variants.  Our analysis suggests that KdV solitons may indeed be formed in the nucleus 
with a width which, in some cases, can be smaller than one fermi. 

\end{abstract} 

\maketitle



\vspace{1cm}
\section{Introduction}

Understanding the nucleus and its properties  in terms of QCD is a very ambitious goal 
and we are still far from it. Significant progress has been achieved in this direction
with the use of  chiral perturbation theory \cite{weise}. 
On the other hand, 
relativistic mean field (RMF) models  have been sucessfully  used to describe the bulk 
properties of nuclear matter, finite nuclei and dense stars \cite{furn,abcm,lala,serot}. 
Although they are only effective models, they are simple, depend only on few parameters and 
are the ideal tool to test certain aspects of nuclear matter. 

In this work we shall use the relativistic mean field model discussed in \cite{lala} to 
investigate the formation and propagation of Korteveg - de Vries solitons in nuclear matter. 
Assuming that nuclear 
matter behaves like an ideal fluid we wish to know whether small baryon density perturbations 
can propagate without dissipation through the medium. If this turns out to be case, these 
pulses  might be responsible for interesting phenomena. One of them would be the apparent
nuclear transparency in proton nucleus collisions at low energies. The impinging proton would 
be absorbed by the fluid and turned into a density pulse. It would then be able to traverse 
the 
target nucleus emerging on the other side. This idea has been first advanced in \cite{frsw}. 
In \cite{rawei} the experimental evidence of this ``nuclear transparency''  was discussed.
In \cite{rww} the formalism was improved with the inclusion of viscosity and 
three-dimensional 
expansion. In \cite{hef} the effect of the nuclear surface was considered. One important 
lesson 
from all these works is that the formation of solitons depends in a crucial way on the 
equation of state (EOS).
In \cite{frsw,rawei,rww,hef} the equation of state  employed was borrowed from 
\cite{glass} and it was very phenomenological, containing no information about the 
microscopic nucleon dynamics.

The experimental signature of the nuclear soliton  is very difficult to 
measure, since the outcoming nucleon emerges in a direction very close to the beam. Probably 
because of this reason the subject was left aside for almost ten years. Soliton 
formation in nuclear matter  was then revived in \cite{abu}, where a different  
equation of state was  considered. This EOS was developed in
\cite{vaut} and  was  based on the nucleon-nucleon microscopic interaction given by 
the Skyrme force.  The main finding of  \cite{abu} was that the nuclear  soliton is not 
a compression pulse, as thought before.  Instead, it is a rarefaction pulse due to a sign 
inversion appearing because of the use of  Skyrme interactions. 

Now, another ten years later, we come back to this subject because we feel that this 
beautiful idea  has not been explored enough. We shall replace the EOS used 
in the early works by another one,  based on the model of \cite{lala}. 
The usual relativistic  mean field  models do not generate equations of state which lead 
to solitons because they do not contain higher derivatives, required in  nonlinear 
differential equations.  Therefore we will  introduce a new interaction term in the 
Lagrangian of   \cite{lala}.

This work is organized as  follows. In the next section we present some simple formulas of 
nonrelativistic hydrodynamics and write them in terms of the baryon and energy density. This 
will help to establish the connection between these equations and the relativistic mean field 
EOS. In the subsequent section we discuss the model of  \cite{lala}, introduce the new 
interaction term and, in the mean field approximation, derive the equation of state. In the 
following section, we combine the hydrodynamic equations with the new EOS and derive the 
KdV equation and its analytical solution. The last section contains some concluding
remarks.

\section{Hydrodynamics}
The basic equations of nonrelativistic hydrodynamics \cite{hidro} are 
Euler equation
\begin{equation}
{\frac{\partial \vec{v}}{\partial t}} +(\vec{v} \cdot \nabla) \vec{v}=
-\bigg({\frac{1}{\rho}}\bigg) \nabla p 
\label{euler1}
\end{equation}
and the continuity equation 
\begin{equation}
{\frac{\partial \rho}{\partial t}} + {{\nabla}} \cdot (\rho {\vec{v}})=0  
\label{cont1}
\end{equation} 
In these equations, $\rho$ is the mass density, which,  in the  nuclear medium, is
related to the baryon density through 
\begin{equation}
\rho=M\rho_{B}
\label{ro}
\end{equation}
where $M$ is the nucleon mass. 
The enthalpy per nucleon is given by \cite{hidro}:
\begin{equation}
dh=Tds+Vdp 
\label{ent1}
\end{equation}
where \hspace{0.2cm} $V=1/\rho_{B}$ \hspace{0.2cm} is the specific volume. 
For a perfect fluid  $(ds=0)$    the  equation above becomes  $dp=\rho_{B}dh$ and 
consequently:
\begin{equation}
 \nabla p=\rho_{B}\nabla h  
\label{gradpe}
\end{equation}
Inserting (\ref{ro}) and  ({\ref{gradpe}}) in ({\ref{euler1}}) and in (\ref{cont1}) 
we obtain  
\begin{equation}
{\frac{\partial \vec{v}}{\partial t}} +(\vec{v} \cdot \nabla) \vec{v}=
-\bigg({\frac{1}{M}}\bigg) \nabla h 
\label{eulerentalpia}
\end{equation}
and
\begin{equation}
{\frac{\partial \rho_B}{\partial t}} + {{\nabla}} \cdot (\rho_B {\vec{v}})=0  
\label{contibari}
\end{equation} 
The enthalpy per nucleon may also be written as \cite{abu}:
\begin{equation}
h=E+\rho_{B}{\frac{\partial E}{\partial\rho_{B}}} 
\label{entalbu}
\end{equation} 
where $E$ is the energy per nucleon given by:
\begin{equation}
E=\frac{\varepsilon}{\rho_{B}}
\label{epb}
\end{equation}
which inserted in ({\ref{entalbu}}) leads to:
\begin{equation}
h={\frac{\partial \varepsilon }{\partial \rho_{B}}} 
\label{entalderiv}
\end{equation}
and therefore:
\begin{equation}
\nabla h= \nabla {\frac{\partial \varepsilon }{\partial \rho_{B}}} 
\label{gradh}
\end{equation} 
With these simple manipulations we can make explicit  the dependence of  
Euler equation (\ref{eulerentalpia})  on $\varepsilon(\rho_{B})$, or in other 
words, on the  equation of state  of  nuclear matter.

As it will be seen later, it is possible to perform a combination of
({\ref{eulerentalpia}}) and ({\ref{contibari}}) in a specific way to obtain a nonlinear 
differential equation
for the baryon density, with soliton  solutions.
The existence or not of these solitons depends on the function $\varepsilon(\rho_{B})$.  
A key feature of ({\ref{eulerentalpia}}) and (\ref{gradh}) is that they contain 
the gradient of the derivative of the energy density. If $\varepsilon$ contains a 
Laplacian of $\rho_B$, i.e., 
\begin{equation}
{\mathcal{\varepsilon}} \propto ... + ... \nabla^{2} \rho_{B} + ...
\label{epslap}
\end{equation}
then equation ({\ref{eulerentalpia}}) will have a cubic derivative
with respect to the space coordinate,
which will give rise to the Korteweg-de Vries equation for the density. 
The most popular  RMF models do not have higher derivative terms and, even if they have at
the start, these terms are usually neglected during the calculations.  
In the next section we show how to overcome these limitations
and obtain a slightly more general expression for  $\varepsilon(\rho_{B})$  with 
Laplacians.

\section{Modified QHD}

We start with  the well known variant of QHD-I \cite{serot}, employed in \cite{lala}:
$$
\mathcal{L}_{0}=\bar{\psi}[\gamma_{\mu}(i \partial^{\mu} - g_{V}V^{\mu})-(M-g_{S} \phi)]\psi 
+ 
{\frac{1}{2}}\Big(\partial_{\mu} \phi \partial^{\mu} \phi - {m_{S}}^{2} \phi^{2}\Big)
$$
\begin{equation}
-{\frac{1}{4}}F_{\mu \nu}F^{\mu \nu} +{\frac{1}{2}}{m_{V}}^{2}V_{\mu}V^{\mu}   
-{\frac{b\phi^{3}}{3}}-{\frac{c\phi^{4}}{4}}
\label{lagratotal}
\end{equation}
where, as usual, the degrees of freedom are 
the baryon field $\psi$, the neutral scalar meson field $\phi$
and the neutral vector meson field $V_{\mu}$, with the respective couplings and masses. 
The parameters $b$ and $c$ will be taken from \cite{lala}. 
As it can be seen, there are no higher order derivatives. This is the reason why 
the traditional QHD and most of its  numerous variants do not give an equation 
of state compatible with soliton propagation, like (\ref{epslap}). For
this  it is necessary, in first place,  to modify the Lagrangian, introducing a term 
with Laplacian derivative couplings.  The simplest one is given by:
\begin{equation}
{\mathcal{L_{M}}} \equiv {\frac{g_{V}}{{m_{V}}^{2}}}\bar{\psi}
(\partial_{\nu} \partial^{\nu} V_{\mu})\gamma^{\mu} \psi 
\label{lagram}
\end{equation}
where the subscrit $\mathcal{M}$ denotes ``modification'' in QHD. First of all, this term 
is designed to be small in comparison with the main baryon - vector meson interaction term
$g_{V} \bar{\psi} \gamma_{\mu} V^{\mu}  \psi$. Because of the derivatives, it is of the 
order of:
\begin{equation}
\frac{p^2}{m_V^2} \sim \frac{k_F^2}{m_V^2} \sim 0.12
\label{estimate}
\end{equation}
where the Fermi momentum is $k_F\simeq 0.28$ GeV and $m_V \simeq 0.8$ GeV. 
The form chosen for ${\mathcal{L_{M}}}$ is not dictated by any symmetry argument, has no 
other deep justification and is just one possible interaction term among many others 
\cite{cdmg}. It is used here as a  prototype. In the future, we plan 
to address this problem with a state of the art Lagrangian.  As it can be immediately 
guessed, the integration by parts of  ${\mathcal{L_{M}}}$ leads to a D'Alembertian 
acting on the baryon current. We present (\ref{lagram}) with the derivatives acting on 
$ V^{\mu}$ just because it can be more easily compared to 
$g_{V} \bar{\psi} \gamma_{\mu} V^{\mu}  \psi$ it allows for estimates like 
(\ref{estimate}). The total Lagrangian is then:
\begin{equation}
{\mathcal{L}}=\mathcal{L}_{0} + \mathcal{L}_{M}
\label{ltotal}
\end{equation}
Integrating  $\mathcal{L}_{M}$ by parts twice,  making the standard mean field 
approximation 
\begin{equation}
V_{\mu} \rightarrow <V_{\mu}> \equiv \delta_{\mu 0} V_{0} 
\label{vzero}
\end{equation}
\begin{equation}
\phi \rightarrow <\phi> \equiv \phi_{0}  
\label{phizero}
\end{equation}
and neglecting the time and spatial derivatives of the fields $V_{0}$ and $\phi_{0}$,
we arrive at the approximate Lagrangian $\mathcal{L'}$ given by:
$$
\mathcal{L'}=\bar{\psi}[(i\gamma_{\mu} \partial^{\mu} - g_{V}\gamma_{0}V_{0})
-(M-g_{S} \phi_{0})]\psi -{\frac{1}{2}}{m_{S}}^{2}{\phi_{0}}^{2}+{\frac{1}{2}}
{m_{V}}^{2}{V_{0}}^{2}
$$
\begin{equation}
-{\frac{b{\phi_{0}}^{3}}{3}}-{\frac{c{\phi_{0}}^{4}}{4}}
+{\frac{g_{V}}{{m_{V}}^{2}}}[\partial_{\nu} \partial^{\nu}
(\bar{\psi} \gamma^{0} \psi )]V_{0}
\label{llinhaexpli}
\end{equation}
The equations of motion of the fields are:
\begin{equation}
[(i\gamma_{\mu} \partial^{\mu} - g_{V}\gamma_{0}V_{0})
-(M-g_{S} \phi_{0})]\psi=0
\label{eqm1}
\end{equation}
\begin{equation}
{{m_{V}}^{2}}V_{0}= g_{V} \bar{\psi} \gamma^{0} \psi \, 
\label{eqm2}
\end{equation}
\begin{equation}
{m_{S}}^{2}\phi_{0} = g_{S} \bar{\psi} \psi -b{\phi_{0}}^{2}-c{\phi_{0}}^{3}
\label{eqm3}
\end{equation}
where, for simplicity, we have neglected the terms containing Laplacians. This means 
that the perturbation changes slightly the energy of the system, but does not affect 
the dynamics of the environment, i.e., of the unperturbed $\psi$, $\phi$ and 
$V^{\mu}$ fields.  The energy density  is calculated using the energy-momentum tensor:
\begin{equation}
\varepsilon =<T_{{0}{0}}>=-g_{00}\mathcal{L'} +
(\partial_{0}\psi){\frac{\partial \mathcal{L'}}{\partial (\partial^{0}
\psi)}}
\label{enerd}
\end{equation}
which, with the help of (\ref{llinhaexpli}), (\ref{eqm1}), (\ref{eqm2}) is given by:
$$
\varepsilon={\frac{{g_{V}}^{2}}{2{m_{V}}^{2}}}\rho_{B}^{2} 
+{\frac{{m_{S}}^{2}}{2}}{\bigg[{\frac{(M^{*}-M)}{g_{S}}}\bigg]}^{2}+
$$
$$
+{\frac{\gamma}{(2\pi)^{3}}}\int_{0}^{k_{F}} d^3{k} ({\vec{k}}^{2}+{M^{*}}^{2})^{1/2}  
+{\frac{b}{3g_S^3}}(M^{*}-M)^{3}
+{\frac{c}{4g_{S}^{4}}}(M^{*}-M)^{4}
$$
\begin{equation}
+ {\frac{{g_{V}}^{2}}
{{m_{V}}^{4}}}\rho_{B}
\nabla^{2}\rho_{B}-{\frac{{g_{V}}^{2}}{{m_{V}}^{4}}}\rho_{B}
\bigg({\frac{\partial^{2}\rho_{B}}{\partial t^{2}}}\bigg)
\label{epsilonexp}
\end{equation}
where $(\gamma)$ is  the baryon  spin-isospin degeneracy factor and   
$M^*$ stands for the nucleon effective mass:
$$
M^{*} \equiv M-g_{S}\phi_{0}
$$
The two terms in  the last line of (\ref{epsilonexp}) come from  ${\mathcal{L_{M}}}$. 
Baryon number propagation in nuclear matter has been studied in \cite{shin} with the help 
of the diffusion equation: 
\begin{equation}
{\frac{\partial \rho_{B}}{\partial t}} = D \, \nabla^{2}\rho_{B}
\label{diffusion}
\end{equation}
where the diffusion constant $D$ was numerically evaluated as a function of density and 
temperature and found to be $D\simeq \, 0.35 \,\, fm$ at 
densities comparable to the equilibrium 
nuclear density and temperatures of the order of $80$ MeV. The exact value of $D$ at 
$T=0$ is unknown but its  dependence both with temperature and density is very smooth 
and an extrapolation to $T=0$ points to values even smaller than $0.35$ fm. This number 
is small compared to any nuclear size scale and can be interpreted  as indicating that 
density gradients do not disappear very rapidly in nucler matter.
Using (\ref{diffusion}) twice in the
last term of (\ref{epsilonexp}) it can be rewritten as:
\begin{equation}
-{\frac{{g_{V}}^{2}}{{m_{V}}^{4}}}\rho_{B}
\bigg({\frac{\partial^{2}\rho_{B}}{\partial t^{2}}}\bigg)=
-{\frac{{g_{V}}^{2}}{{m_{V}}^{4}}}\rho_{B}{\frac{\partial}{\partial t}}\bigg(D 
\nabla^{2}\rho_{B}\bigg) = 
-{\frac{{g_{V}}^{2}}{{m_{V}}^{4}}}\rho_{B}D^{2}[\nabla^{2}(\nabla^{2}\rho_{B})]
\label{sectimederi}
\end{equation}
which, in the context of the present calculation, can be neglected  because
$\nabla^{2}(\nabla^{2}\rho_{B})<<(\nabla^{2}\rho_{B})$. 
With this last approximation, the final form of the energy density is given by 
(\ref{epsilonexp}) without the last term. The baryon density up to the Fermi level 
$(k_{F})$ is given by:
$$
\psi^{\dagger} \psi \equiv \rho_{B}={\frac{\gamma ({{k_{F}})^{3}}}{6 \pi^{2}}}
$$
As usual, the  nucleon effective mass  is determined by the minimization  of 
$\varepsilon(M^{*})$ with respect to $M^{*}$, which results in  the self-consistency 
relation:
$$
M^{*}=M-{\frac{{g_{S}}^{2}}{{m_{S}}^{2}}}{\frac{\gamma}{(2 \pi)^{3}}} 
\int^{k_{F}}_{0} d^3{k} 
{\frac{M^{*}}{\Big({\vec{k}}^{2}+{M^{*}}^{2}\Big)^{1/2}}} 
$$
$$
-{\frac{g_S^{2}}{ms^{2}}}\bigg[ {\frac{b}{g_S^{3}}}(M^{*}-M)^{2} + 
{\frac{c}{g_S^{4}}}(M^{*}-M)^{3} 
\bigg]
$$

The main result of this section is (\ref{epsilonexp}), which, in what follows, will be 
used in the derivation of the KdV equation. By keeping $\nabla^{2}\rho_{B}$ (but not 
$\nabla^{2} \nabla^{2}\rho_{B}$) we take into account the small inhomogeneities, which 
do not change rapidly in time 
(and therefore we neglect $\frac{\partial \rho_{B}}{\partial t}$).
Whereas this approximation is quite often used at the nuclear surface, it is not very 
common for the bulk of nuclear matter. We believe, however, that it is quite reasonable.


\section{The KdV  solitons}

In this section we follow the treatment developed in \cite{frsw} and \cite{abu}
to obtain the Korteweg-de Vries  equation in one dimension by the combination 
of (\ref{eulerentalpia}) and (\ref{contibari}).  

Expression  (\ref{epsilonexp}) is valid for any $\rho_B$. In order to make contact with
the nuclear matter phenomenology and use the constraints imposed by the saturation curve, 
we will now use  (\ref{epsilonexp}) to compute the energy per baryon $E$ (\ref{epb}). 
We then Taylor expand it around the equilibrium density $\rho_{0}$ up to second order:
\begin{equation}
E(\rho_{B})=E(\rho_{0})
+\frac{1}{2}  
\bigg({\frac{\partial^{2} E}{{\partial\rho_{B}}^{2}}}\bigg)_{\rho_{B}=\rho_{0}} 
(\rho_{B}-\rho_{0})^{2} 
\label{energypernucleon}
\end{equation}
where  the first order term vanishes because of the saturation condition
$$
{\frac{\partial }{{\partial\rho_{B}}}}\bigg({\frac{\varepsilon}
{\rho_{B}}}-M\bigg)_{\rho_{B}=\rho_{0}}\hspace{0.5cm}=
\hspace{0.5cm}\bigg({\frac{\partial E}{{\partial\rho_{B}}}}\bigg)_{\rho_{B}=\rho_{0}}
\hspace{0.5cm}=\hspace{0.5cm}0
$$
The derivative coefficient in (\ref{energypernucleon}) may be rewritten as \cite{abu}:
$$
\bigg({\frac{\partial^{2} E}{{\partial\rho_{B}}^{2}}}\bigg)_{\rho_{B}=\rho_{0}} \, = \,
\frac{M{c_{s}}^{2}}{\rho_{0}^{2}}
$$
where $M$ is the nucleon mass and $c_{s}$ is the speed of sound in nuclear matter. 
Calculating the enthalpy per nucleon (\ref{entalbu}) by using (\ref{energypernucleon}) 
leads to: 
\begin{equation}
h=E(\rho_{0})+{\frac{M{c_{s}}^{2}}{2{\rho_{0}}^{2}}}(3{\rho_{B}}^{2}
+{\rho_{0}}^{2}-4{\rho_{B}}{\rho_{0}}) 
\label{entalpypernucleon}
\end{equation}
where
\begin{equation}
E(\rho_{0})=\bigg({\frac{{g_{V}}^{2}}{2{m_{V}}^{2}}}\bigg){\rho_{0}}
+ {\frac{\eta(\rho_{0})}{\rho_{0}}}
+\bigg({\frac{{g_{V}}^{2}}{{m_{V}}^{4}}}\bigg)(\nabla^{2}\rho_{B}) 
\label{44}
\end{equation}
and
$$
\eta(\rho_{0}) \equiv {\frac{{m_{S}}^{2}}{2}}{\bigg[{\frac{(M^{*}-M)}{g_{S}}}\bigg]}^{2}
+{\frac{\gamma}{(2\pi)^{3}}}\int_{0}^{k_{F}} d^3{k} ({\vec{k}}^{2}+{M^{*}}^{2})^{1/2} +
$$
$$ 
+{\frac{b}{3g_S^{3}}}(M^{*}-M)^{3}
+{\frac{c}{4g_S^{4}}}(M^{*}-M)^{4}
$$
where the integration limit $k_F$ (which appears also in $M^*=M^*(k_F)$)  
is fixed at the value corresponding to $\rho_{0}$.
Inserting  (\ref{entalpypernucleon}) in (\ref{eulerentalpia}) we arrive at:
$$
{\frac{\partial \vec{v}}{\partial t}} +(\vec{v} \cdot \nabla) \vec{v}=
-\bigg({\frac{{c_{s}}^{2}}{2 {\rho_{0}}^{2}}}\bigg)[6{\rho_{B}}(\nabla \rho_{B})
-4{\rho_{0}}(\nabla{\rho_{B}})] 
$$
\begin{equation}
-\bigg({\frac{{g_{V}}^{2}}{M{m_{V}}^{4}}}\bigg)\nabla(\nabla^{2}\rho_{B}) 
\label{eulerfinal}
\end{equation}
We now repeat the steps developed in \cite{frsw} and \cite{abu},  
write  (\ref{eulerfinal}) and  (\ref{contibari}) in one space dimension ($x$) and
introduce dimensionless variables for the density and velocity: 
\begin{equation}
\hat\rho={\frac{\rho_{B}}{\rho_{0}}}
\label{rochapeu}
\end{equation}
and
\begin{equation}
\hat v={\frac{v}{c_{s}}} 
\label{vchapeu}
\end{equation}
We next  define the ``stretched coordinates''  $\xi$ and $\tau$ as in 
\cite{frsw,abu,davidson}:
\begin{equation}
\xi=\sigma^{1/2}{\frac{(x-{c_{s}}t)}{R}} 
\label{xi}       
\end{equation}
and
\begin{equation}
\tau=\sigma^{3/2}{\frac{{c_{s}}t}{R}} 
\label{tau}       
\end{equation}
where $R$ is a size scale and $\sigma$ is a small ($0 < \sigma < 1$) expansion parameter 
chosen to be \cite{davidson}:
\begin{equation}
{\sigma} = {\frac{\mid u-{c_{s}} \mid}{{c_{s}}}}
\label{sigma}       
\end{equation}
where $u$ is the propagation speed of the perturbation. After these tedious but 
straightforward substitutions, we expand (\ref{rochapeu}) and (\ref{vchapeu}) around 
their equilibrium values:
\begin{equation}
\hat\rho=1+\sigma \rho_{1}+ \sigma^{2} \rho_{2}+ \dots
\label{roexp}
\end{equation}
\begin{equation}
\hat v=\sigma v_{1}+ \sigma^{2} v_{2}+ \dots
\label{vexp}
\end{equation}
Equations (\ref{eulerfinal}) and (\ref{contibari}) will contain power series in $\sigma$  
up to $\sigma^2$. Since the coefficients in these series are independent we get a set of 
equations, which, when combined, lead to the KdV equation for $\rho_{1}$
\begin{equation}
{\frac{\partial {\rho}_{1}}{\partial \tau}}+
3{{\rho}_{1}}{\frac{\partial{\rho}_{1}}{\partial \xi}}
+\bigg({\frac{{g_{V}}^{2}{\rho_{0}}}{2M{c_{s}}^{2}{m_{V}}^{4}R^{2}}}\bigg)
{\frac{\partial^{3}{\rho}_{1}}{\partial \xi^{3}}}=0  
\label{KdV}
\end{equation}
which has a well known soliton solution. We may rewrite the above equation back in  
the $x \, - \, t$ space obtaining a KdV-like equation for ${\hat{\rho}_{1}}$ with the 
following solitonic solution:
\begin{equation}
{\hat{\rho}_{1}}(x,t)={\frac{(u-{c_{s}})}{{c_{s}}}}
sech^{2}\bigg[
{\frac{{m_{V}}^{2}}{{g_{V}}}}\sqrt{{\frac{(u-{c_{s}}){c_{s}}M}
{2{\rho_{0}}}}}(x-ut) \bigg] 
\label{solchapeu}
\end{equation}
where ${\hat{\rho}_{1}} \equiv \sigma {\rho_{1}}$.
The solution above is a bump with width $\lambda$ given by:
\begin{equation}
\lambda={\frac{g_{V}}{{m_{V}}^{2}}} \sqrt{{\frac{2{\rho_{0}}}{(u-{c_{s}}){c_{s}}M}}}
\label{width}
\end{equation}
Since 
$$ 
0 < sech^{2} (\theta) \leq 1
$$
and since  
\begin{equation}
{\frac{u-c_{s}}{c_{s}}} = \sigma  \leq 1
\label{ucs}
\end{equation}
we can see that ${\hat{\rho}_{1}}$ is always smaller than one and is
consistent with the expansion (\ref{roexp}). Relation (\ref{ucs}) implies that  
 $u\leq u_{max}=2 c_{s}$.  This upper bound is arbitrary, being a consequence of
the choice (\ref{sigma}), but it must exist. As mentioned in the introduction, the 
perturbation of the medium may be caused by the passage of a nucleon projectile and
we  expect that for higher projectile energies the modification of medium will be 
so strong that our perturbative approach will break down. The upper bound in the pulse
velocity implies a limit for the  momentum of the incident nucleon given by:
\begin{equation}
p_{max}={\frac{2c_{s}M}{{\sqrt{1-4{c_{s}}^{2}}}}} 
\label{pmax}
\end{equation}

As it can be seen from (\ref{width}), the soliton width depends on parameters of 
the microscopic dynamics, 
as the coupling and mass which come from the Lagrangian (\ref{ltotal}). The sound  
velocity comes from the EOS derived from the same Lagrangian. The expression (\ref{width})  
also reveals the supersonic nature of this phenomenon, since it is only possible 
for $u > c_s$.   

In the case where the density pulse is caused by a nucleon traversing a nucleus, if the 
width is comparable to the nucleon size and much smaller than the nuclear radius, 
the pulse will be localized and the impinging nucleon may emerge from the nucleus. If, 
on the other hand, $\lambda$ is large, of the order of the nuclear radius, then the 
incident energy will most likely be absorbed by the nucleus as a whole, which will  
undergo some collective excitation and decay.

In the numerical evaluation of (\ref{width}) we need to know the values of the
parameters appearing in the Lagrangians (\ref{lagratotal}) and  (\ref{lagram}). 
For $\mathcal{L}_{0}$ we consider both the original QHD-I, without the nonlinear terms, 
just as a reference, and the most phenomenologically successful  model of 
\cite{lala}, with its different parameter choices labeled by  $NL1$, $NL3$, $NL3-II$ and 
$NL-SH$. The use of the original  QHD-I equation of state in hydrodynamical calculations 
leads to a too rapidly expanding system, as it was  shown in \cite{mnno}. 

All the input numbers as well as the calculated  incompressibilities $K$, 
equilibrium densities $\rho_0$, nucleon effective masses $M^*$ and 
sound velocities $c_s$ are shown in Table 1. 
\begin{center}
\begin{tabular}{|l|l|l|l|l|l|r|}  \hline  \hline  
         & $QHD$ & $NL1$ & $NL3$   & $NL3-II$ & $NL-SH$       \\ \hline  
K(MeV)   & 545       & 211   & 272  & 272  & 355        \\ \hline
M(MeV)   & 939       & 938   & 939     & 939      & 939           \\ \hline
$m_{S}(MeV)$   & 500   & 492  & 508,2 & 507,7 & 526    \\ \hline
$m_{V}(MeV)$   & 780   & 783 & 782,5 & 781,9 & 783            \\ \hline
$g_{S}$   & 8,7   & 10,14  & 10,22 & 10,2 & 10,4   \\ \hline
$g_{V}$   & 11,62  & 13,28 & 12,87 & 12,8 & 12,9    \\ \hline
$M^{*}/{M}$ & 0,56   & 0,57  & 0,6 & 0,59 & 0,6   \\ \hline
${\rho_{0}}(fm^{-3})$ & 0,19   & 0,15  & 0,15 & 0,15 & 0,15   \\ \hline
$b(fm^{-1})$ & 0   & -12,17  & -10,43 & -10,4 & -6,9  \\ \hline
$c$  & 0 & -36,26 & -28,9 & -28,9 & -15,8  \\ \hline
$c_{s}$ & 0,25   & 0,16 & 0,18 & 0,18 & 0,2   \\ \hline \hline  
\end{tabular}
\end{center}
\begin{center}
\small {\bf Table 1: input parameters  and results for $K$, $M^{*}$, $\rho_{0}$ 
and $c_{s}$.} 
\end{center}
It is interesting to observe that the maximal momentum $p_{max}$ depends on 
$c_s$ and therefore will be different for different models. 

Choosing values for $u$, from $c_s$ to $u_{max}$, and computing both $\lambda$ and 
$p$ up to $p_{max}$ for the different models, we obtain the curves shown in Fig. 1. 
As it can be seen, the width decreases with the projectile energy, 
as anticipated in \cite{raha}.  
We can also see that, the stiffer is the equation of state, the wider is the window for 
soliton formation.  At bombarding energies of a few hundred MeV the collisions are 
probably too violent and the departure from the initial equilibrium state is too
large for the soliton picture to be true.  At energies smaller than $200$ MeV, the
width starts to be larger but still significantly smaller than the nuclear radius. 
Moreover, the choice of $g_V$ in (\ref{lagram}) was just a first guess. The derivative 
coupling could be different and much weaker, leading to narrower solitons.  As it appears  
now in (\ref{width}) $\lambda$  is directly proportional to $g_V$. If the coupling would be 
half of the value used here, even the softest EOS ($NL1$) would lead to narrow solitons. 
After having established the connection between KdV solitons and relativistic mean field 
models, we can test other theories and other derivative couplings.

The observation of KdV solitons in nucleon-nucleus collisions was also discussed in 
\cite{clare}, considering the  exisiting experimental results at that time.  Nowadays much 
more data are available and it might be interesting to look at them with the soliton 
picture in mind, searching for some unexpected transparency.


\section{Summary and conclusions}

Assuming that RMF models give a satisfactory description of nuclear matter and finite 
nuclei and, also, that the nucleus can be treated as a perfect fluid, we have investigated 
the possibility of  formation and propagation of Korteweg - de Vries solitons in nuclear 
matter. We have updated the early estimates of the soliton properties, using more  reliable 
equations of state. In doing so we had to introduce in the Lagrangian a term with derivative 
coupling between the baryon and the vector meson fields.  The term used here was the simplest   
one and with it we could already  find KdV solitons, which have their 
properties (in particular the width) controled by the
microscopic parameters of the underlying theory, such as masses and couplings.
Whereas the quantitative results are not entirely conclusive, they suggest that KdV may 
indeed exist in nuclear matter. It is fair to say that the subject deserves further 
investigation.
Finally we would like to mention that, during the last decade, due to the CERN-SPS and RHIC 
programs,  hydrodynamics has experienced a great revival \cite{heinz,hama} and supersonic 
phenomena \cite{super} were also  brought back to explain recent RHIC data. Nonlinear effects, 
as those discussed here, may be relevant also in this context.

\begin{acknowledgments}
We wish to express our gratitude to S. Raha for numerous suggestions and useful 
comments and hints. We are also most grateful to M. Malheiro, M. Chiapparini,  
S.S. Avancini and 
D. Bazeia for enlightening discussions. This work was  partially 
financed by the Brazilian funding
agencies CAPES, CNPq and FAPESP. 
\end{acknowledgments}




\begin{figure} \label{fig1}
\centerline{\psfig{figure=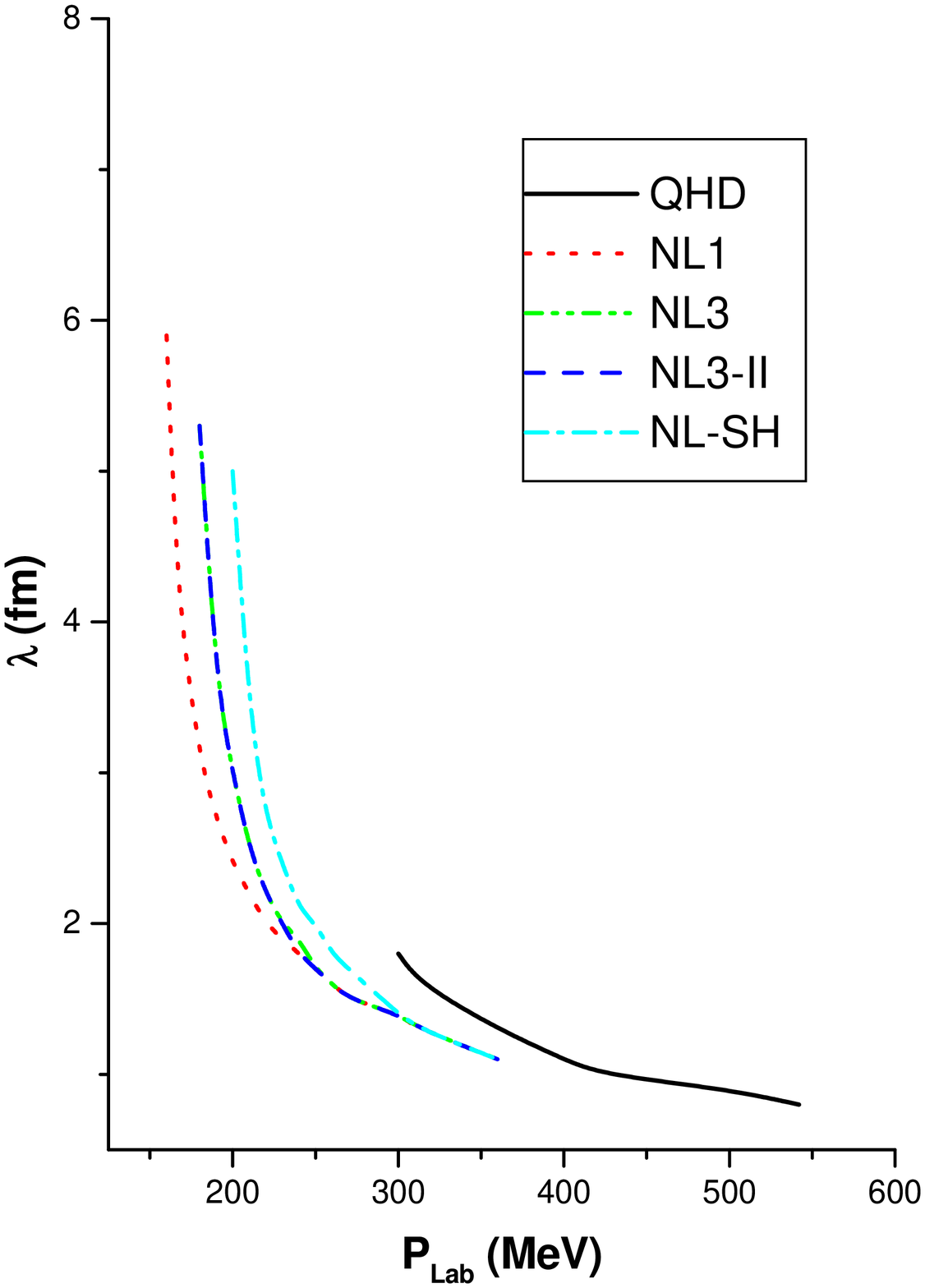,width=16cm,angle=0}}
\vspace{-1.5cm}
\caption{Soliton width as a function of the projectile momentum}
\end{figure}

\end{document}